\documentclass[a4paper, 10pt, conference]{ieeeconf}      % Use this line for a4 paper

\IEEEoverridecommandlockouts                              % This command is only needed if 
\usepackage{graphics} % for pdf, bitmapped graphics files
\usepackage{epsfig} % for postscript graphics files
\usepackage{mathptmx} % assumes new font selection scheme installed
\usepackage{times} % assumes new font selection scheme installed
\usepackage{amsmath,amsfonts} % assumes amsmath package installed
\usepackage{amssymb}  % assumes amsmath package installed
\usepackage{array}
\usepackage{textcomp}
\usepackage{bm}
\usepackage{algorithm}
\usepackage{algorithmic}
\usepackage{threeparttable}
\usepackage{booktabs}

\title{\LARGE \bf
Centralized Coordination of Connected Vehicles at Intersections\\
using Graphical Mixed Integer Optimization*
}

\author{Qiang Ge, Qi Sun, Zhen Wang, Shengbo Eben Li, Ziqing Gu and Sifa Zheng% <-this % stops a space
\thanks{*This work is supported by National Key R\&D Program in China with 2018YFB1600600.}% <-this % stops a space
\thanks{Q. Ge, Q. Sun, Z. Wang, S. Li, Z. Gu and S. Zheng are with the State Key Laboratory of Automotive Safety and Energy, Tsinghua University, Beijing, 100084, China. All correspondence should be sent to S. Zheng (Email: zsf@tsinghua.edu.cn)}%
}

\begin{document}

\maketitle
\thispagestyle{empty}
\pagestyle{empty}

%%%%%%%%%%%%%%%%%%%%%%%%%%%%%%%%%%%%%%%%%%%%%%%%%%%%%%%%%%%%%%%%%%%%%%%%%%%%%%%%
\begin{abstract}
This paper proposes a centralized multi-vehicle coordination scheme serving unsignalized intersections. The whole process consists of three stages: a) target velocity optimization: formulate the collision-free vehicle coordination as a Mixed Integer Linear Programming (MILP) problem, with each incoming lane representing an independent variable; b) dynamic vehicle selection: build a directed graph with result of the optimization, and reserve only some of the vehicle nodes to coordinate by applying a subset extraction algorithm; c) synchronous velocity profile planning: bridge the gap between current speed and optimal velocity in a synchronous manner. The problem size is essentially bounded by number of lanes instead of vehicles. Thus the optimization process is real-time with guaranteed solution quality. Simulation has verified efficiency and real-time performance of the scheme.

\end{abstract}

%%%%%%%%%%%%%%%%%%%%%%%%%%%%%%%%%%%%%%%%%%%%%%%%%%%%%%%%%%%%%%%%%%%%%%%%%%%%%%%%
\section{Introduction}

Intersection capacity is a limit to transportation efficiency. A jammed crossing will deteriorate safety, efficiency, gas emission, as well as passengers’ experience (because of frequently stop-and-go operations). Much effort has been contributed to this field due to fore-mentioned reasons.

Researches on vehicles at signalized intersections mostly focus on the energy consumption \cite{c-1}\cite{c0}, while there are different concerns about the coordination of vehicles at intersections without traffic signal—the balance between safety and efficiency, the real-time performance and quality of solution, as well as the dynamic change of variables to be optimized. 

The classic principle of “First Come, First Serve” (FCFS) was proposed in \cite{c1}, where vehicles dynamically apply for right of way to the intersection agent, and the latter responds according to current reservations by previous vehicles. However, no much optimization is done with regard to passing sequence in such reservation-based methods.

Optimization-based intersection management is an active field, where the numerical solution has been an issue. Li \textit{et al} \cite{c2} propose a concept of safety driving patterns to represent the collision-free movements of vehicles at crossings. As the authors point out, the complexity of this cooperative driving planning will increase quickly with the number of vehicles. Kamal \textit{et al} \cite{c3} adopt a risk expression based on a 2-D Gaussian function, of which the variables are vehicles’ distance to the intersection. The proposed approach requires initial values or guess of the solution. Joyoung Lee \textit{et al} \cite{c4} develop a Cooperative Vehicle Intersection Control (CVIC) system that enables cooperation between fully automated vehicles and infrastructure. By eliminating the potential overlaps of vehicular trajectories, the CVIC algorithm seeks a safe maneuver for every vehicle. The system performs well in simulation, but there may exist infeasible solutions, for which the authors have designed recovery modes.

Among these optimization methods, mixed integer linear programming is relatively popular for intersection management \cite{c5}\cite{c6}\cite{c7}. By introducing binary variables as a mathematical expression for different passing sequences, arrangement of sequence can be transformed into a MILP formulation. The real-time requirement may not be satisfied because of computation delay \cite{c5}. Additionally, velocity of vehicles are not optimized globally.

Since the objective function consists of items representing each vehicle, it may dynamically change because of the entering/leaving of vehicles, probably before the solution is done. Obviously, to guarantee the real-time performance as well as quality of the solution, and adapt to the objective function’s dynamic changing feature, decomposition is key to the intersection vehicle coordination problem. One way to decompose is to control vehicles in a distributed manner, which will decrease the communication demand and reduce the size of optimal problem, and no central agent is needed. Researches using virtual platoon \cite{c8}, model predictive distributed control \cite{c9}, multi-agent reinforcement learning \cite{c10}, and the alternating direction method of multipliers (ADMM) \cite{c10.5} have been introduced. These works have preconditions or limits, such as centralized problem construction at certain steps, rule-based relative priority, or the lack of interpretability.

Oriented Graph is usually adopted to describe relative priorities between vehicles \cite{c8}\cite{c11}\cite{c12}. Specially, Yu Chao \textit{et al} \cite{c11} used Coordination Graph (CG) to decompose the global payoff function   into linear combination of local payoff functions. To realize a multiagent coordinated learning effect, they apply the Variable Elimination (VE) algorithm \cite{c13}. The VE algorithm sequentially eliminates agent nodes till only one remains, and then the last agent selects the action that maximizes the payoff function. Similarly, in this paper we solve the optimal velocity of vehicles after nodes in a graph is eliminated to minimum.

In this paper, we propose a centralized coordination scheme, where the MILP formulation deals with the safety and efficiency simultaneously. Scale of the optimization problem is limited, by introducing a subset extraction method based on graph theory. Thus, the real-time performance of solution is improved, and the scheme can be easily applied into practice since it does not require the group of vehicles to be fixed.

The remnant of this paper is arranged as follows. In Section II, a multi-vehicle passing problem is built with all due assumptions and simplifications. Section III presents the methodology of centralized vehicle coordination. The simulation and results are illustrated in Section IV. Section V gives some concluding remarks.

\section{Problem Formulation}

\begin{figure}[htbp]
    \centerline{\includegraphics[width=5.97cm,height=4.43cm]{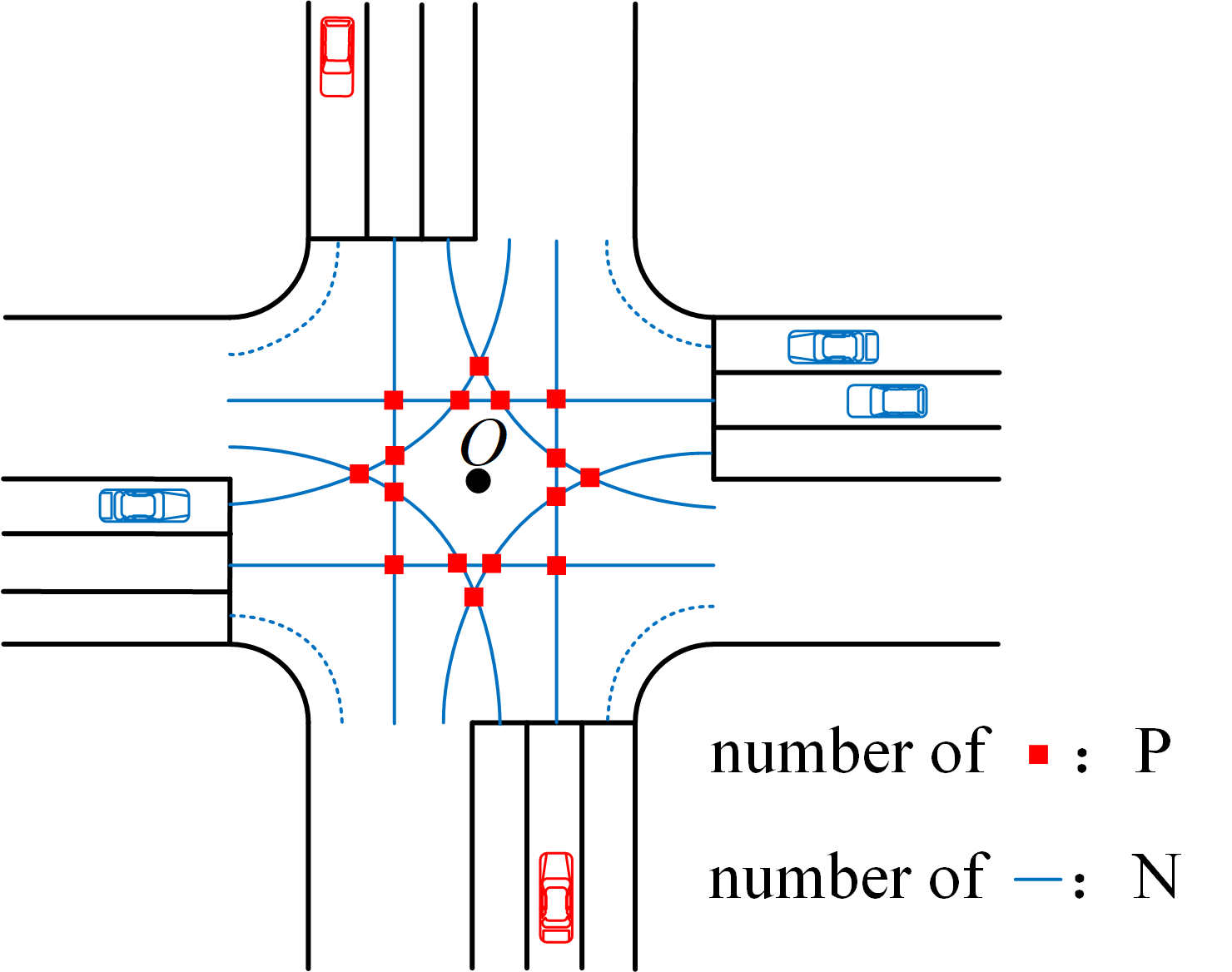}}
    \caption{Problem formulation: intersection scenario}
\label{FIG1-a}
\end{figure}

This paper chooses a multi-lane intersection scenario as a general example shown in Fig. \ref{FIG1-a}, and we suppose there are enough lanes so that vehicles will not merge. Thus the form of conflict between all paths is only crossing. Vehicles turning right can accelerate/deaccelerate freely. Denote number of all the incoming traffic flows to be coordinated as $N$, and number of all crossing points as $P$.

\label{problem}

Another assumption holds that vehicles are already on their target lanes and their trajectories are all predefined, thus only longitudinal advisory speed is considered. Communication condition is ideal without any latency or package loss. Vehicles approach the intersection with different but numerically close initial velocity. Without losing generality, their departure time can be stochastic, as long as keeping reasonable headway.

For simplicity of demonstration, take 4-leg single-lane intersection as an example, where vehicles only go straightly. This example differs only in scale from Fig. \ref{FIG1-a} essentially, and the latter is the scenario applied in simulation.  

\section{Methodology}

\subsection{Concepts and Notations} For the $i$th vehicle, denote its distance to intersection center $O$ as $L_i$. In following section, an approach is introduced to dynamically coordinate different sets of vehicle, each time vehicles constituting the set are chosen from the $N$ lanes. Denote set of their distances as ${\textbf{L}}_{1 \times N}$. Define matrix ${\textbf{C}}_{N \times N}$, if $i \in N$, $j \in N$ have their paths crossing, then $c_{ij}=1$, otherwise the element is set to be 0.

At each crossing point, there exists a relative priority between two vehicles. Define ${\textbf{S}}_{N \times N}$, where
\begin{equation}
    \begin{array}{l}
        {s_{ij}}: = \left\{ \begin{array}{l}
        1,{\rm \ vehicle}\ i\ {\rm has\ priority\ over\ vehicle }\ j\\
        -1,{\rm \ vehicle }\ j\ {\rm has\ priority\ over\ vehicle }\ i\\
        0,{\ {\rm paths\ of\ vehicle}\ i\ {\rm and\ vehicle\ }j\ {\rm do\ not\ cross}}
        \end{array} \right.
        \end{array}
\end{equation}

According to definition in last section we know that there are $P$ positive values and $P$ negative values (i.e. 1 and -1) in ${\textbf{S}}_{N \times N}$.
% Build a directed graph $\textbf{G}$=\{$\textbf{L}$,$\textbf{S}$\}, and use $s_{ij}$ to indicate the direction of edge. 
Table \uppercase\expandafter{\romannumeral1} gives important expressions in this paper and their meanings.

\begin{table}
    \caption{Notations and Meanings}
    \label{table1} 
    \begin{center}
    \begin{tabular}{p{1.5cm}<{\centering} p{6cm}}
    \hline
    \textbf{Symbol}&\textbf{Description}\\
    \hline
    {$N$} & number of incoming lanes coordinated \\

    {$N'$} & number of elements in the extracted subset \\
 
    {$P$} & number of crossing points between $N$ paths\\

    %{$W$} & width of the intersection \\
    
    {$L_i$} & distance between vehicle $i$ and intersection center \\
    
    {${\textbf{L}}_{1 \times N}$} & $\{L_i|1\leq i \leq N \}$ \\
    
    {${\textbf{C}}_{N \times N}$} & conflicting matrix, $c_{ij}=1$ if movements of $i$ and $j$ cross, otherwise $c_{ij}=0$ \\
    
    {${\textbf{S}}_{N \times N}$} & priority matrix, $s_{ij}=1$ if $i$ is prior than $j$, $s_{ij}=-1$ if $j$ is prior than $i$, $s_{ij}=0$ if $i$, $j$ are irrelevant\\
    
    {$v_{opt,i}$} & optimal velocity for vehicle $i$\\
    
    {${\textbf{V}}_{1 \times N}$} & $[v_{opt,1},\cdots,v_{opt,i},\cdots,v_{opt,N}]^T$\\

    {${\textbf{V}}_{1 \times N'}$} & optimal velocity vector of the subset\\

    {$v_{0,i}$} & initial velocity of vehicle $i$\\

    {${\textbf{V0}}_{1 \times N}$} & $[v_{0,1},\cdots,v_{0,i},\cdots,v_{0,N}]^T$\\

    {${\textbf{V0}}_{1 \times N'}$} & initial velocity vector of the subset\\
    
    {${\textbf{G}}_{1 \times N}$} & directed graph, with ${\textbf{V}}_{1 \times N}$ stored in vertexes, and all $s_{ij} \neq 0$ indicating directed edges\\
    \hline
    %\hline
    \end{tabular}
    \end{center}
\end{table}

\subsection{2-vehicle collision-free constraint}

\begin{figure}[htbp]
    \centerline{\includegraphics[width=6.5cm,height=4cm]{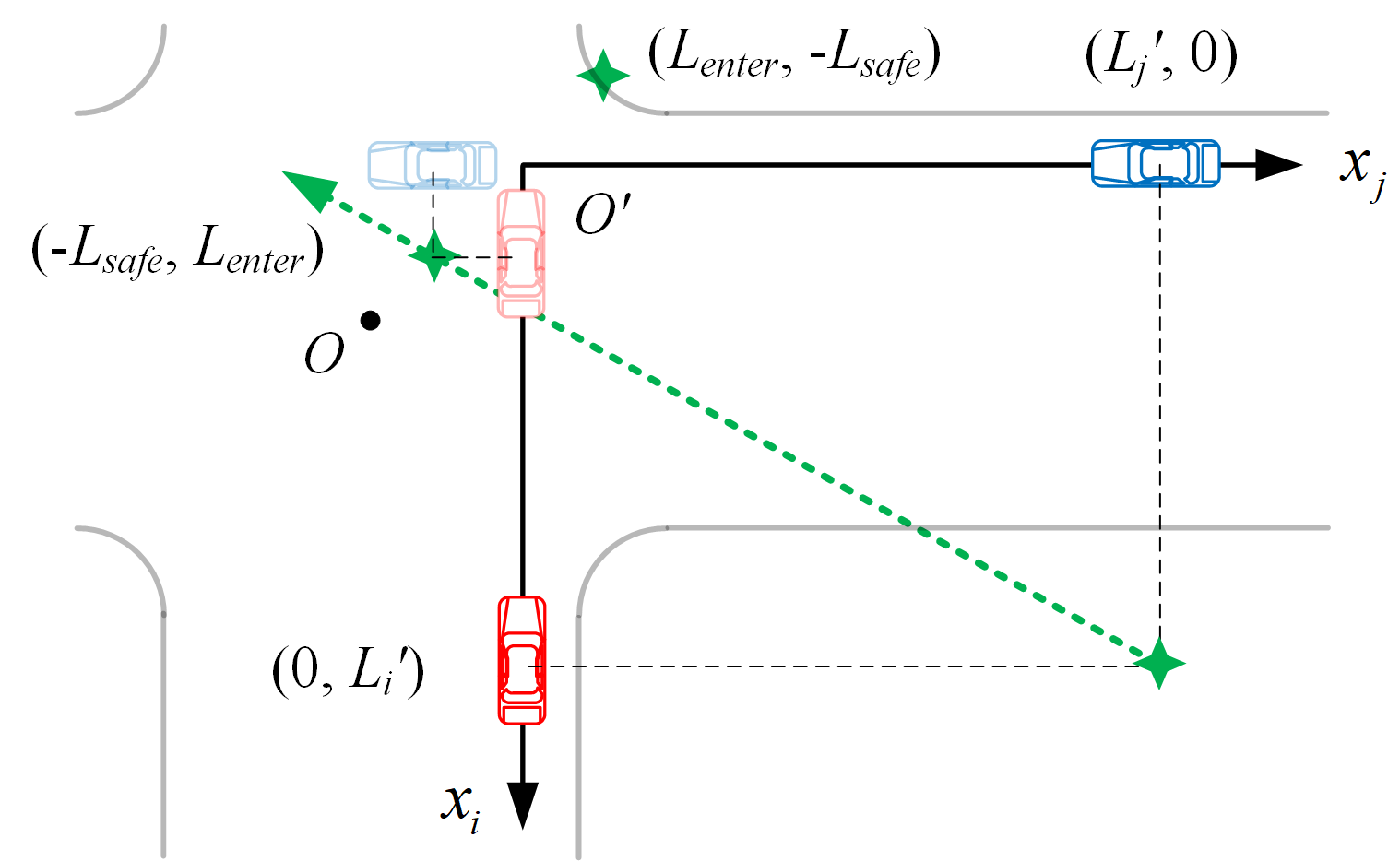}}
    \caption{Collision-free joint velocity model}
\label{FIG1-b}
\end{figure}
% Predefined trajectories could have two forms of conflict: merging, crossing. Correspondingly, vehicles on different paths may have side collision with others.
For $\forall$ $(i, j)$ with $c_{ij}=1$, build a Cartesian coordinate system as shown in Fig. 2. If one or both of the paths are curves, map their trajectories onto the axes. 
Since two vehicles share a potential conflict point $O'$ on their intended trajectories, to guarantee safety, never should they appear at the point simultaneously. This constraint can be expressed in geometry by defining a joint trajectory (green dotted line in Fig. \ref{FIG1-b}), of which transverse coordinate is $x_j$, and longitudinal coordinate is $x_i$. The joint trajectory should either go under $(-L_{safe},L_{enter})$, or beyond $(L_{enter},-L_{safe})$. The two points represent vehicle $j$ leaves before/after vehicle $i$ respectively. Here $L_{enter}$ refers to distance from centroid of vehicle to origin $O$ when the vehicle head just touches the overlapped region. Similarly, $L_{safe}$ refers to distance limit from centroid of the other vehicle to origin $O$ at the same time. Both values can be adjusted to change safety margin. Notice that the origin $O'$ is biased with regard to intersection center $O$, we define ${L_i}'$ and ${L_j}'$, based on coordination transformation. Obviously the feasible region has deterministic linear boundary if joint trajectory is a straight line. Therefore, we suppose that%the speed ratio $v_i / v_j$ remains constant with regard to time, making it easy to describe feasible region in mathematical form.

% Similarly, intersections of different types can be categorized by the amount and layout of conflict points. 

\newtheorem{assumption}{\textbf{Assumption}}
\begin{assumption}
    Both vehicles drive at constant speed from current position to the intersection area, without acceleration or deceleration.
\label{AS1}
\end{assumption}

Given initial locations ${L_i}'$ and ${L_j}'$ in local coordinate system, we have
\begin{equation}\frac{{{L_i}' - {x_i}(t)}}{{{v_i}}} = t = \frac{{{L_j}' - {x_j}(t)}}{{{v_j}}}\end{equation}
Thus following equation is obtained:
\begin{equation}
    {v_i}{x_j}(t) - {v_j}{x_i}(t) = {L_j}'{v_i} - {L_i}'{v_j}
    \label{EQ3}
\end{equation}
The basic 2-vehicle collision constraint can be converted into inequality expressions:
\begin{equation}
    \begin{array}{l}
        {x_i}(t) \le -L_{safe},{x_j}(t) = L_{enter}\\
            \qquad\qquad\qquad\vee \\
        {x_j}(t) \le -L_{safe},{x_i}(t) = L_{enter}
    \end{array}
\label{EQ4}
\end{equation}
From (\ref{EQ4}), it is apparent that the feasible region of joint trajectory line is determined by three points on the coordinate plane, i.e. $(-L_{safe},L_{enter})$, $(L_{enter},-L_{safe})$, and start point $(L_j,L_i)$, which is in accordance with Fig. \ref{FIG1-b}.
Combining (\ref{EQ3}) and (\ref{EQ4}), we get
\begin{equation}
    \begin{array}{l}
        (L_{enter} - {L_j}'){v_i} \le ( - L_{safe} - {L_i}'){v_j}\\
        \qquad\qquad\qquad\qquad\vee \\
        (L_{enter} - {L_i}'){v_j} \le ( - L_{safe} - {L_j}'){v_i}
    \end{array}
\label{EQ5}
\end{equation}
Expression (\ref{EQ5}) cannot be used in an optimal problem directly, except that the inequalities have a logical relation of AND, “$\wedge$”. Hence, the big-M method is applied here \cite{c6}\cite{c14}. Introduce a binary variable $b_{ij}$ and a relatively big constant value $M$, and (\ref{EQ5}) now equals
\begin{equation}
    \begin{array}{l}
        (L_{enter}-{L_j}'){v_i}-(1-{b_{ij}})M\le(-L_{safe}-{L_i}'){v_j}\\
        \qquad\qquad\qquad\qquad\qquad\wedge \\
        (L_{enter}-{L_i}'){v_j}-{b_{ij}}M \le(-L_{safe}-{L_j}'){v_i}\\
        \\
        {\rm subject\ to}\qquad\qquad {b_{ij}} \in \{{0,1}\}
    \end{array}
\label{EQ6}
\end{equation}

So far, under \textbf{Assumption \ref{AS1}} the 2-vehicle collision avoidance is formulated as inequality constraints, with variables $v_i$, $v_j$ and $b_{ij}$ to be optimized.

\subsection{Basic MILP Problem}
\label{basic MILP}

From each approaching direction pick one vehicle and build an linear programming problem: $\max \sum\limits_{i = 1}^N {{v_i}}$, subject to ${v_i} \in [{v_{min}},{v_{max}}]$. 

The feasible region is further constrained by (\ref{EQ6}). Note that $b_{ij}$ is either 1 or 0, making it a MILP problem.
If there is no vehicle in incoming lane $i$ currently, $L_i$ can be set infinity. Then ${v_{opt,i}}$ will definitely be ${v_{max}}$, and other variables are totally not affected.

\newtheorem{remark}{\textbf{Remark}}
\begin{remark}
By now, based on \textbf{Assumption \ref{AS1}}, given current position of certain vehicles (i.e. one from each approaching lane), solution to the MILP problem is optimal passing velocity ${\{v_{opt,i}\}}$ and their priorities ${\{b_{ij}\}}$. Here ${\{v_{opt,i}\}}$ is denoted as ${\textbf{V}}_{1 \times N}$, and $b_{ij}$ determines ${\textbf{S}}_{N \times N}$. 
% The objective of optimization is maximum average passing velocity of certain vehicle group (each from one movement), which improves the traffic efficiency indirectly.
\label{RK1}
\end{remark}

\subsection{Intersection coordination decomposed into sub-problems}
\label{break into sub}

We propose a coordination scheme consisting of 3 stages, which are introduced below.
\subsubsection{Target Velocity Optimization}\label{AA}

Given vehicles to be coordinated at current stage, the MILP problem is solved with their positions as input.
Consider following question: 
\begin{equation}
    \label{MILP formulation}
\begin{array}{l}
    \max \sum\limits_{i=1}^N {{v_i}} \\
    {\rm s.t.}\ A \bm{x} \le \bm{b} \\
    \qquad {v_i} \in [{v_{min}},{v_{max}}] \\
    \qquad {b_{ij}} \in \{0,1\}
    \end{array}
\end{equation}

where
\begin{equation}
    \begin{array}{l}
        A = \left[ {\begin{array}{*{20}{c}}
        {{A_1}}&{{A_2}}
        \end{array}} \right] = {\left[ {\begin{array}{*{20}{c}}
        {{A_{11}}}&{}\\
        \vdots &{}\\
        {{A_{1k}}}&{{A_2}}\\
        \vdots &{}\\
        {{A_{1P}}}&{}
        \end{array}} \right]_{2P \times (N + P)}}
    \end{array}
\end{equation}

\begin{equation}
    \begin{array}{l}
        {A_{1k}} = {\left[ {O\begin{array}{*{20}{c}}
            {{L_{enter}} - {L_j}'}\\
            {{L_{safe}} + {L_j}'}
            \end{array}O{\rm{ }}\begin{array}{*{20}{c}}
            {{L_{safe}} + {L_i}'}\\
            {{L_{enter}} - {L_i}'}
            \end{array}{O}} \right]_{2 \times N}}\\
                \begin{small}
                \begin{array}{*{20}{c}}
                    \qquad\quad \cdots &\  col.\ i\quad& \cdots &\quad col.\ j& \cdots 
                \end{array}
                \end{small}
    \end{array}
    \label{EQ12}
\end{equation}

\begin{equation}
    \begin{array}{l}
        {A_2} = {\left[ {\begin{array}{*{20}{c}}
            M&{}&{}\\
            { - M}&{}&{}\\
            {}& \ddots &{}\\
            {}&{}&M\\
            {}&{}&{ - M}
            \end{array}} \right]_{2P \times P}}
    \end{array}
\end{equation}

\begin{equation}
    \begin{array}{l}
        {\bm{x}} = \left[ {\begin{array}{*{20}{c}}
            {{v_1}}& \cdots &{{v_N},}& \cdots &{{b_{ij}}}& \cdots 
            \end{array}} \right]_{1 \times (N + P)}^T
    \end{array}
    \label{EQ14}
\end{equation}

\begin{equation}
    \begin{array}{l}
         \bm{b} = \left[ {\begin{array}{*{20}{c}}
            M&0&M&0& \cdots &M&0
            \end{array}} \right]_{2P}^T
    \end{array}
\end{equation}

Input variables are ${\textbf{L}}_{1 \times N}$ and ${\textbf{C}}_{N \times N}$. ${\textbf{L}}_{1 \times N}$ determines values in (\ref{EQ12}) after simple coordination transformation ($O$ to $O'$). And indexes of non-zero values in ${\textbf{C}}_{N \times N}$ determines all $(i,j)$ pairs in $A_{11},\cdots,A_{1P}$.

To make the algorithm concise, it does not involve vehicle size explicitly. Actually, vehicle lengths and widths are used as additional input parameters of the function, and they finally influence the result by changing $L_{safe}$ and $L_{enter}$ in (\ref{EQ6}) and (\ref{EQ12}). 

Output of the MILP optimization is ${\textbf{V}}_{1 \times N}$ and the $P$ values of $b_{ij}$.

\subsubsection{Vehicle Dynamic Selection}

\begin{figure}[htbp]
    \centerline{\includegraphics[width=8cm,height=4.20cm]{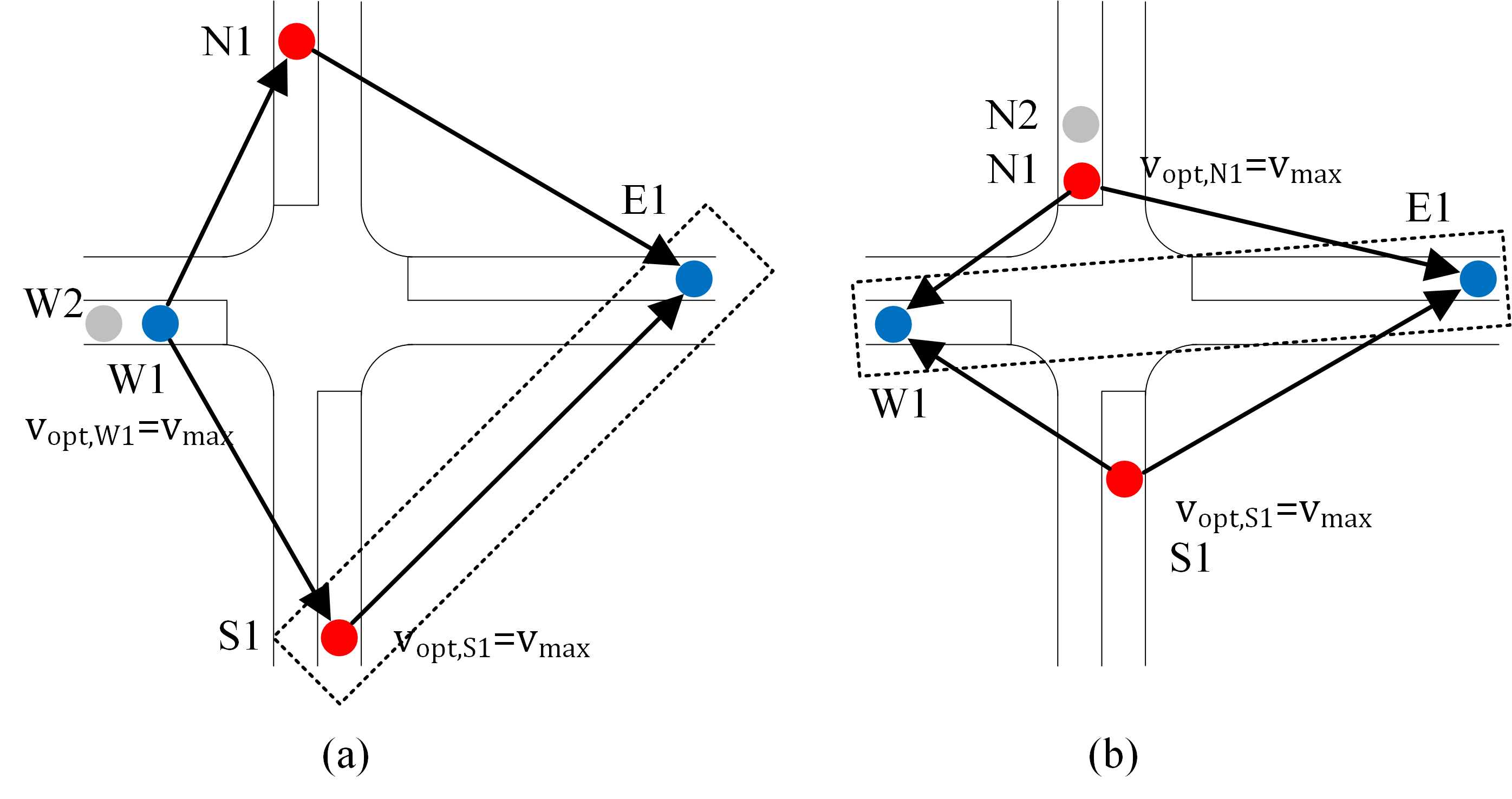}}
    \caption{Excluding node demonstration}
    \label{FIG7}
\end{figure}

The basic idea is to choose one vehicle from each lane of $N$ to form a subset. In Fig. \ref{FIG7} vehicles are represented by nodes. Grey nodes mean the vehicle is either not coordinated yet or already coordinated. The colored nodes represent vehicles chosen into current coordination subset, with different traffic phases expressed by red or blue. No matter how many vehicles are driving into the intersection, each running time the MILP problem only need to deal with $N$ nodes, namely \{E1,N1,W1,S1\} in Fig. \ref{FIG7}.

Here we make use of the directed graph ${\textbf{G}}_{1 \times N}$ to find excluded nodes. Calculate the optimal velocity and sequence of $N$ vehicles by constructing a basic MILP problem described in \textit{1)}. The optimal velocity is then denoted as ${\textbf{V}}_{1 \times N}$, and the relative passing priority which comes from $b_{ij}$ in (\ref{EQ14}) is stored in ${\textbf{S}}_{N \times N}$. With $N$ vehicles nodes and their relative priorities it is easy to build a directed graph ${\textbf{G}}_{1 \times N}$. 
Vehicle node $k$ is excluded if and only if one of the items below is satisfied:

\begin{enumerate}
    \item optimal velocity assigned to the $k$th node is $v_{max}$, indegree of node $k$ is not zero, and node $k$ is not an element in a loop
    \item there exist two nodes $i$ and $j$, both assigned $v_{max}$ and having an indegree of zero, while node $k$ is their common child node
    \item node $k$ is child node of any node satisfying 1) or 2)
\end{enumerate}

In the directed graph, the edge pointing from $j$th node to $i$th node means vehicle $j$ is prior to vehicle $i$, and vice versa. Let's call a vehicle assigned $v_{max}$ free node, and others constrained node. The reason behind above three items is intuitive.

In a directed acyclic graph (DAG), optimized velocity equalling upper limit implicitly means the ego vehicle is either the first to pass the intersection, or following some other vehicles much closer than it. Otherwise it should drive slower than speed limit. The latter case, reflected in the graph, will be at least one directed edge pointing to the node, namely its indegree is $\geq 1$, while the former case keeps its indegree zero. In this sense, a free node with indegree $\geq 1$ should be expelled because it can drag down the overall speed. As is shown in Fig. \ref{FIG7}(a), node W2 has to follow node S1, even it is closer to stop line, which is not allowed. The only exception of this rule rises when the graph is not DAG and the free node is on a circuit. In this case, the none-zero indegree value does not stand for a much higher distance to the intersection, so the node should be reserved in the subset. So far, the rule 1) is concluded.

For constrained nodes, the constrained optimal velocity reflects that they need to drive slower to avoid crash. These nodes should be reserved, because they are closely coupled with predecessors instead of following them with a long headway. But there still can be an exception: suppose there are two free nodes in parallel phases, indegree of both are zeros, thus both not excluded according to rule 1). In Fig. \ref{FIG7}(b), they are N1 and S1. However, remaining distances of the two free nodes can be distinct, making the child node of the farther one arriving even later, like E1 following S1 in Fig. \ref{FIG7}(b). In this case, N2 would have to wait until E1 leaves, which is not reasonable enough. Therefore, E1 should be excluded and optimized again in next round. 

Rule 3) is easy to understand: once a node is excluded to enhance overall efficiency, nodes forming its spanning tree should be expelled for the same reason.

\begin{remark}
    Excluding nodes avoids unnecessary deacceleration, especially when traffic flow density is unbalanced in different phases. By this means, our decomposition of intersection coordination problem is essentially adaptive.
    \label{RK2}
\end{remark}

The algorithm is as following:

\begin{algorithm}[htb]
  \renewcommand{\algorithmicrequire}{\textbf{Input:}}
  \renewcommand{\algorithmicensure}{\textbf{Output:}}
 \caption{Subset Extraction Algorithm}
 \label{algorithm1}
 \begin{algorithmic}
  \REQUIRE ${\textbf{V}}_{1 \times N}$, ${\textbf{S}}_{N \times N}$
  \ENSURE  $FLAG_{1 \times N}$
  \STATE $FLAG_{1 \times N}:=[1\ 1\ \cdots 1]_{1 \times N}$
  
  \STATE build directed graph ${\textbf{G}}_{1 \times N}$ := \{${\textbf{V}}_{1 \times N}$, ${\textbf{S}}_{N \times N}$\}
  \STATE $Leading Set := \{ i|v_{opt,i} = v_{max}\}$

  \FORALL{$i \in Leading Set$}
  \IF{indegree($i$) $\neq 0 $}
  \STATE $FatherSet_i \Leftarrow $predecessors$(i) $
  \IF{$FatherSet_i \not\subset $spanningtree$(i) $}
  \STATE $FLAG($spanningtree$(i)) \Leftarrow 0 $
  \ENDIF
  \ENDIF
  \ENDFOR
  
  \FORALL{$i,j\in Leading Set, i\neq j, i\not\in$neighbor($j$)}
  \IF{indegree($i$) = indegree($j$) = $0$}
  \STATE $Intersect_{ij} \Leftarrow$ spanningtree$(i) \cap $spanningtree$(j)$
  \IF{$Intersect_{ij} \neq \emptyset$}
  \STATE $FLAG($spanningtree($Intersect_{ij}$)$) \Leftarrow 0$
  \ENDIF
  \ENDIF
  \ENDFOR

 \end{algorithmic}
\end{algorithm}

The spanningtree(), neighbor(), predecessors() are graph theory involving functions, which return a set of nodes. indegree() is also a function which returns number of edges pointing to its input node.

The result of \textbf{Algorithm} \ref{algorithm1} is $FLAG_{1 \times N}$, and $FLAG(i)=0$ means the $i$th node should be excluded. We denote the size of the extracted subset as $N'$.

\subsubsection{Synchronous Velocity Profile Planning}
\begin{figure}[htbp]
      \centerline{\includegraphics[width=7cm,height=10.43cm]{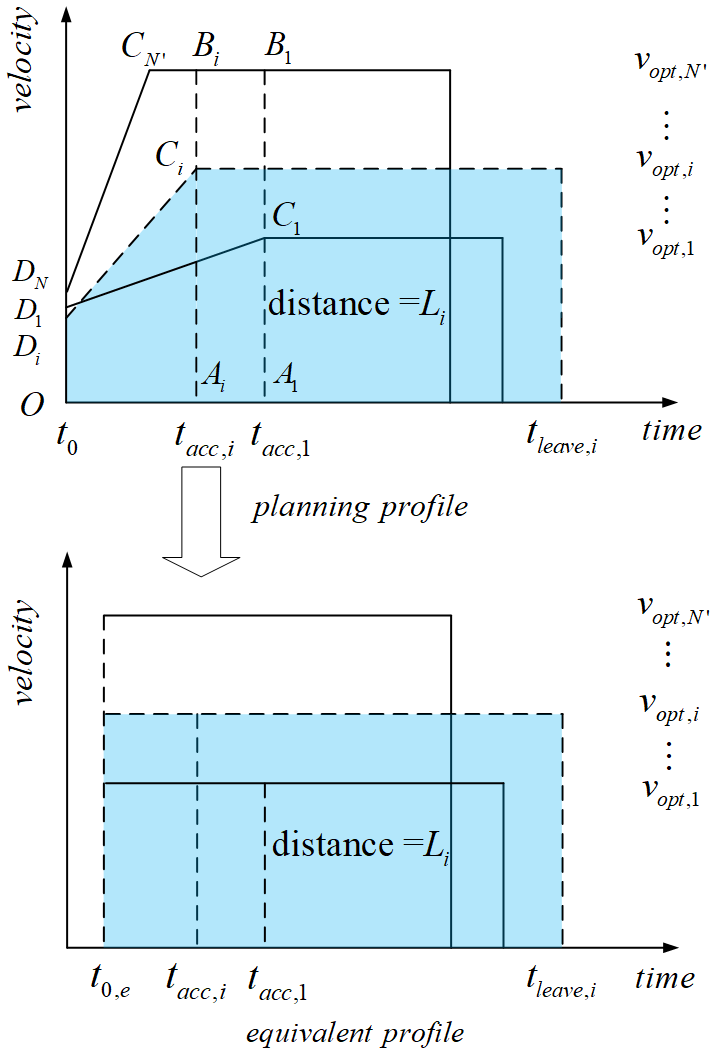}}
      \caption{Synchronous velocity profile planning}
  \label{FIG3}
  \end{figure}

% Note that according to \textbf{Assumption \ref{AS1}}, vehicles are supposed to drive at their own $v_{opt}$ respectively, all the way from their initial position to the intersection area. 
Actually, \textbf{Assumption \ref{AS1}} is a relaxation to practical constraint, because velocity must be continuous with regard to time due to vehicle dynamics. Therefore, we must ensure that, the process of vehicles accelerating/deaccelerating, is equivalent to a period they drive at their respective optimal velocity (Fig. \ref{FIG3}). By this means, \textbf{Assumption \ref{AS1}} is violated only in transient process ($t_0 \sim t_{acc,i}$ in Fig. \ref{FIG3}), optimal solution of $N$ vehicles' MILP problem at $t_0$ is consistent with new solution after max($t_{acc,i}$). $t_{acc,i}$ means the time when the $i$th vehicle's velocity converges to its target. The conditions to be met are concluded as follows.

\textbf{Synchronous Condition}: By all vehicles' velocity convergence, distances they have travelled since MILP solution should be proportional to their optimal velocity.

\textbf{Transfer Condition}: Vehicles should never collide with members in last coordination subset.

To realize the 2 conditions, a velocity profile planning algorithm is proposed. Here the input contains the subset index $FLAG$, as well as the reduced optimal velocity, initial velocity \textit{et al}. $A_{max}$ is the maximum acceleration of the vehicles, and $K$ is a constant greater than 1. Expressions like $S_{OA_{i}C_{i}D_{i}}$ refer to area illustrated in Fig. \ref{FIG3}. The ${\textbf{t}}_{lastleave,1 \times N}$ comes from last run of this algorithm, namely the output ${\textbf{t}}_{leave,1 \times N}$ generated last time.

\begin{algorithm}[htb]
    \renewcommand{\algorithmicrequire}{\textbf{Input:}}
    \renewcommand{\algorithmicensure}{\textbf{Output:}}
   \caption{Velocity Profile Planning Algorithm}
   \label{algorithm2}
   \begin{algorithmic}
    
    \REQUIRE ${\textbf{V}}_{1 \times N'}$, ${\textbf{V0}}_{1 \times N'}$, ${\textbf{L}}_{1 \times N'}$, ${\textbf{t}}_{lastleave,1 \times N}$, $T_{clock}$, $A_{max}$, $K$, ${\textbf{C}}_{N \times N}$, $FLAG$
    \ENSURE  ${\textbf{t}}_{acc,1 \times N'}$, ${\textbf{t}}_{leave,1 \times N}$
    
    \STATE $N'=$length$({\textbf{V}})$
    \STATE ${\textbf{t}}_{arrive,1 \times N}:=[Inf\ Inf\ \cdots Inf]_{1 \times N}$
    \STATE sort ${\textbf{V}}$ and ${\textbf{V0}}$ to keep ${\textbf{V}(i)}/{\textbf{V0}(i)}$ in ascending order
    \STATE ${\textbf{t}}_{acc}(N') \Leftarrow\displaystyle\frac{{\textbf{V}(N')}-{\textbf{V0}(N')}}{A_{max}}$

    \FOR{$i=1$ to $N'-1$}
        \STATE calculate ${\textbf{t}}_{acc}(i)$ according to $\displaystyle\frac{S_{OA_{i}C_{i}D_{i}}} {S_{OA_{i}B_{i}C_{N'}D_{N'}}}= \frac {v_{opt,i}} {v_{opt,N'} }$
    \ENDFOR

    \STATE update ${\textbf{t}}_{arrive}$ according to $FLAG$

    %\IF {$({t_{arrive,i}})+T_{clock} > t_{lastleave,j}, i,j$ crosses}
    \WHILE{exist $(i,j)$ s.t. ${\textbf{t}}_{arrive}(i) \leq {\textbf{t}}_{lastleave}(j), c_{i,j}=1$}
        \STATE ${\textbf{t}}_{acc} \Leftarrow {\textbf{t}}_{acc} \times K$
        \STATE update ${\textbf{t}}_{arrive}$ according to $FLAG$, ${\textbf{t}}_{acc}$
    \ENDWHILE
    %\ENDIF

    \STATE update ${\textbf{t}}_{leave}$, according to $FLAG$, ${\textbf{t}}_{arrive}$
  
   \end{algorithmic}
  \end{algorithm}

  The \textbf{Synchronous Condition} is guaranteed by a distance-equivalent method, which can also be realized by optimal control. As for the \textbf{Transfer Condition}, latest leaving time of current subset is transferred to the next subset as a constraint of acceleration planning. If vehicles are entering the intersection too early, their velocity profile is rescheduled.

  To be emphasized, the pseudo code does not embody coordinate transformation, neither the size of vehicle body, which should be considered during implementation.
  
  By 3 steps above, a minimum sub-problem is defined and solved. Each sub-problem will deal with collision-free constraint among a subset of vehicles at time $T_{clock}$, as well as safe time interval between subsets. The whole process is chain-like, with computation divided in time domain.

\section{Simulation and Results}

Simulation of 32 vehicles, four in each movement was carried out. Depart time at a distance of 200m is shown in table \ref{tab:dep time}. Here “ES” stands for “from east, into south”, and others are similarly defined.

The simulation was carried out on a computer with an intel i7 CPU, programmed and solved with matlab 2017b.

\begin{table}[tp]

    \centering
    %\fontsize{6.5}{8}\selectfont
    \begin{threeparttable}
    \caption{depart time from 8 approaching directions}
    \label{tab:dep time}
      \begin{tabular}{cccccccc}
      \toprule
      \multicolumn{8}{c}{Departure Time(s)}\cr
      \cmidrule(lr){1-2} \cmidrule(lr){3-4} \cmidrule(lr){5-6} \cmidrule(lr){7-8}
      $ES$&$EW$&$NE$&$NS$&$WN$&$WE$&$SW$&$SN$\cr
      \midrule
      1&2&2&1&2&1&1&2\cr
      6&7&5&4&7&6&6&7\cr
      10&11&8&7&12&9&10&11\cr
      15&16&14&12&17&14&16&15\cr
      \bottomrule
      \end{tabular}
      \end{threeparttable}
  \end{table}

% \subsection{Safety}

% All the trajectories of 32 vehicles in an 80-meter vicinity of the intersection is shown in Fig. \ref{FIG8}, where the horizon plane corresponds to spatial position and the vertical axis is calibrated by time.

% The method explicitly guarantees security by hard constraints in (\ref{MILP formulation}), and we take vehicle trajectories of phase $WE$ as an instance to illustrate. The trajectories of $WE$ phase and four contradicting phases are shown in a 40-meter vicinity of the intersection, where the horizon plane corresponds to spatial position and the vertical axis is calibrated by time. Since for each $WE$ curve, the auxiliary vertical lines go through it without meeting a red curve at the same time, there exist no collision for vehicles in $WE$ phase during the simulation.

% \begin{figure}[htbp]
%     \centerline{\includegraphics[width=8cm,height=4.63cm]{fig6.png}}
%     \caption{Spatio-temporal trajectory of all vehicles}
%     \label{FIG8}
% \end{figure}

% \begin{figure}[htbp]
%     \centerline{\includegraphics[width=8cm,height=6.67cm]{fig7.png}}
%     \caption{Spatio-temporal trajectory of phase $WE$}
%     \label{FIG8-0}
% \end{figure}

\subsection{Efficiency}
\begin{figure}[htbp]
    \centerline{\includegraphics[width=8cm,height=4.89cm]{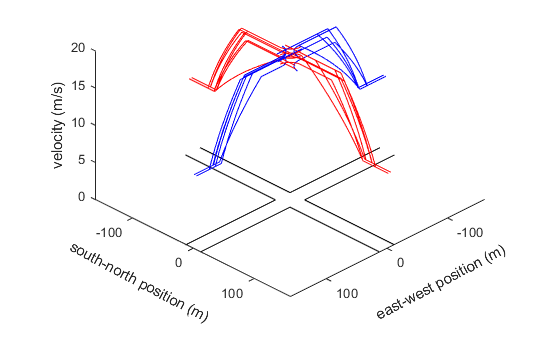}}
    \caption{Velocity trajectory of all vehicles}
    \label{FIG9}
\end{figure}

Velocity trajectories are shown in Fig. \ref{FIG9}, where each curve converges to a vicinity of maximum limit, which is set to be 20 m/s here.

\begin{figure}[htbp]
    \centerline{\includegraphics[width=6cm,height=4.16cm]{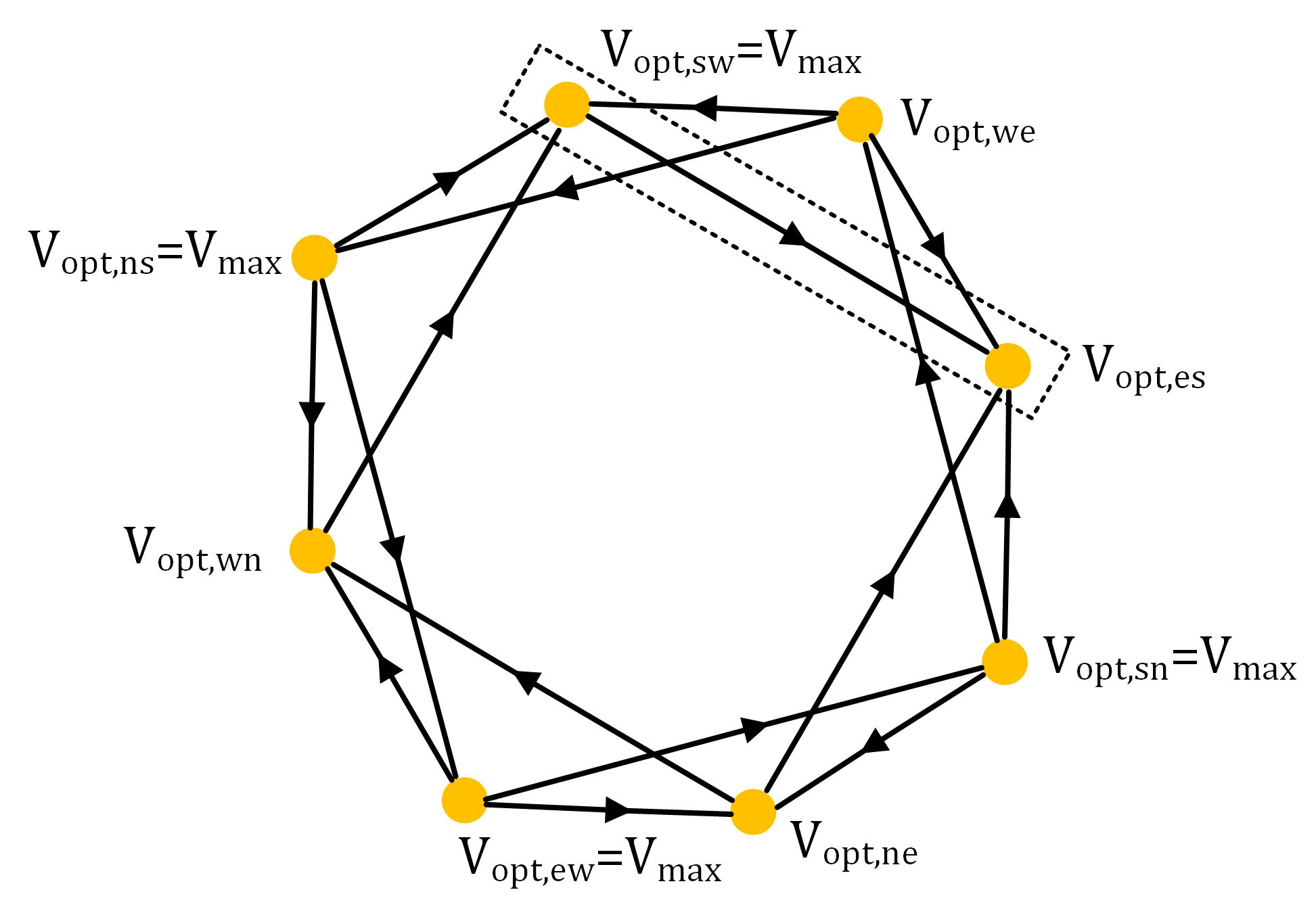}}
    \caption{Directed graph resulting in a 6-vehicle coordination}
    \label{FIG10}
\end{figure}

During the simulation the scheme worked for 16 times, and the subset size is 1, 2, 1, 1, 2, 1, 1, 1, 2, 6, 4, 3, 1, 2, 2, 2 respectively. From the analysis we know, the scheme resembles the FCFS principle when our subset consists of only 1 vehicle. Thus it was 6 that best reveals the essential difference. The directed graph of the topology is given by Fig. \ref{FIG10}. According to node-exclusion rules, only node SW and node ES are excluded, and the left 6 vehicles can accelerate to their optimal velocities and drive through. It is worth mentioning that the four vehicles EN, SN, WE, NS forms a cycle in the topology (Fig. \ref{FIG10}), thus they passes the intersection end to end, as if there exists an invisible roundabout in the intersection at that time.

The time by all the vehicles leave intersection is 31.9 s. We have not compared data strictly with other methods yet, but the above coordinating behavior reveals its potential in improving intersection capacity.
  
\subsection{Real-time Performance}
The mixed integer linear programming accounts for a major part of the computation. The time consumed for the MILP operation during simulation is 0.065, 0.060, 0.057, 0.059, 0.043, 0.041, 0.044, 0.041, 0.046, 0.047, 0.049, 0.053, 0.043, 0.052, 0.045, 0.048 s respectively. The mean value is 0.0495 s and the squared difference is $5.5\times 10^{-5}  s^{2}$. Since the simulation step is chosen to be 0.1 s, the proposed scheme is real-time and feasible.

\section{Conclusion and Discussion}

In this paper, we propose a centralized scheme serving multi-lane intersections. The scheme consists of three steps---target velocity optimization, vehicle dynamic selection and synchronous velocity profile planning. On the premise of safety, the traffic efficiency is improved by optimization of velocity and sequence at the same time. More importantly, real-time solution is satisfied with guaranteed optimality, as the MILP has limited number of variables.

% Based on a MILP formulation, the collision-free management of vehicles from conflicting movements are finished. After that, from all vehicles being managed, we select a subset of them to execute coordination, so as to make the scheme adaptive to continuous and varying traffic flow. Finally, velocity profile of the selected vehicles are planned and transmitted wirelessly. Simulation results reveal that the scheme is real-time and efficient, on the premise of safety.

These aspects should be addressed in future research: design control experiments to test the statistical performance of this scheme, especially compared to other methods; human-driven vehicles should be modelled and considered in the scheme; how to distribute the computation and communication to get rid of the dependence on a central agent.

\addtolength{\textheight}{-12cm}   % This command serves to balance the column lengths

%%%%%%%%%%%%%%%%%%%%%%%%%%%%%%%%%%%%%%%%%%%%%%%%%%%%%%%%%%%%%%%%%%%%%%%%%%%%%%%%

%%%%%%%%%%%%%%%%%%%%%%%%%%%%%%%%%%%%%%%%%%%%%%%%%%%%%%%%%%%%%%%%%%%%%%%%%%%%%%%%

%%%%%%%%%%%%%%%%%%%%%%%%%%%%%%%%%%%%%%%%%%%%%%%%%%%%%%%%%%%%%%%%%%%%%%%%%%%%%%%%

%%%%%%%%%%%%%%%%%%%%%%%%%%%%%%%%%%%%%%%%%%%%%%%%%%%%%%%%%%%%%%%%%%%%%%%%%%%%%%%%

\end{document}